\documentclass[final]{svjour3}

\usepackage{newtxtext,newtxmath}

\usepackage{graphicx,xcolor}
\usepackage{rotating}
\usepackage[numbers]{natbib}

\usepackage[final, colorlinks=true, urlcolor=blue, citecolor=red]{hyperref}

\newcommand{\red}[1]{\textcolor{red}{#1}}

\usepackage{siunitx}
\DeclareSIUnit\sq{\ensuremath\Box}

\makeatletter
\journalname{Journal of Low Temperature Physics}

\bibpunct{[}{]}{,}{n}{}{,}

\begin{document}

\newcommand{\hdblarrow}{H\makebox[0.9ex][l]{$\downdownarrows$}-}
\title{Design of SPT-SLIM focal plane; a spectroscopic imaging array for the South Pole Telescope}

\author{P.~S.~Barry~$^{1,2,3,4}$ \and A.~Anderson~$^{5}$ \and B.~Benson~$^{5}$ \and J.~E.~Carlstrom~$^{1,2,3,6}$ \and T.~Cecil~$^{1}$ \and C.~Chang~$^{1,2,3}$ \and M.~Dobbs~$^{7}$ \and M.~Hollister~$^5$ \and K.~S.~Karkare~$^{2,3}$ \and G.~K.~Keating~$^8$ \and D.~Marrone~$^{9}$ \and J.~McMahon~$^{2,3,6}$ \and J. Montgomery~$^{7}$ \and Z.~Pan~$^1$ \and G.~Robson~$^3$ \and M.~Rouble~$^7$ \and E.~Shirokoff~$^{2,3}$ \and G.~Smecher~$^{10}$}
\authorrunning{P.~S.~Barry et al.}
\institute{
    \email{pbarry@anl.gov}\\
  $^1$ High Energy Physics Division, Argonne National Laboratory, Lemont, IL 60439, USA\\
  $^2$ Kavli Institute for Cosmological Physics, University of Chicago, Chicago, IL, 60637, USA\\
  $^3$ Department of Astronomy and Astrophysics, University of Chicago, Chicago, IL, 60637, USA\\
  $^4$ School of Physics \& Astronomy, Cardiff University, Cardiff, CF24 3AA, UK\\
  $^5$ Fermi National Accelerator Laboratory, Batavia, IL, 60510, USA\\
  $^6$ Department of Physics, University of Chicago, Chicago, IL, 60637, USA\\
  $^7$ Department of Physics and McGill Space Institute, McGill University, Montreal, Quebec H3A 2T8, Canada\\
  $^8$ Harvard-Smithsonian Center for Astrophysics, Cambridge, MA 02138, USA\\
  $^9$ Steward Observatory, University of Arizona, Tucson, AZ 85721, USA\\
  $^{10}$ Three-Speed Logic, Inc., Victoria, B.C., V8S 3Z5, Canada
}

\maketitle

\begin{abstract}

The Summertime Line Intensity Mapper (SLIM) is a mm-wave line-intensity mapping (mm-LIM) experiment for the South Pole Telescope (SPT). The goal of SPT-SLIM is to serve as a technical and scientific pathfinder for the demonstration of the suitability and in-field performance of multi-pixel superconducting filterbank spectrometers for future mm-LIM experiments. Scheduled to deploy in the 2023-24 austral summer, the SPT-SLIM focal plane will include 18 dual-polarization pixels, each coupled to an $R = \lambda / \Delta \lambda = 300$ thin-film microstrip filterbank spectrometer that spans the 2 mm atmospheric window (120-180 GHz). Each individual spectral channel feeds a microstrip-coupled lumped-element kinetic inductance detector, which provides the highly multiplexed readout for the 10k detectors needed for SPT-SLIM. Here we present an overview of the preliminary design of key aspects of the SPT-SLIM the focal plane array, a description of the detector architecture and predicted performance, and initial test results that will be used to inform the final design of the SPT-SLIM spectrometer array.


\end{abstract}

\vspace{-12pt}
\section{Introduction}

Line intensity mapping at mm/sub-mm wavelengths (mm-LIM) is expected to provide a new and complementary probe of large-scale structure (LSS). Using the various emission lines that fall within this wavelength range to trace the underlying matter distribution -- e.g., carbon monoxide rotational ladder (CO), singly ionised carbon (C\textsc{ii}, \SI{158}{\um}) or  nitrogen (N\textsc{ii}, \SI{205}{\um}), and doubly ionised oxygen (O\textsc{iii}, 52 and \SI{88}{\um}) -- mm-LIM is set to provide unique insights into the evolution of LSS out to $z\sim10$. The first generation of experiments dedicated to mm-LIM are well underway, with a recent group of on-sky deployments~\cite{concerto20,time13} along with several new instruments currently under construction \cite{fyst20,tim19}. Most of these experiments are built around technologies that are well established, but are challenging to scale. To efficiently carry out wide-field high-redshift LIM surveys, future mm-LIM experiments will require a new class of large-format spectroscopic imaging arrays.

Superconducting filterbank spectrometers have emerged as a promising candidate for future spectroscopic focal planes operating at mm-wavelengths (70-500 GHz). The integrated on-chip architecture enables a substantial reduction in size and cost relative to existing approaches, allowing multiple spectrometers to be patterned onto a single silicon wafer. Recent progress in spectrometer design and thin-film fabrication techniques has resulted in a number of successful small-scale laboratory demonstrations~\cite{uspec16,camels15,sspec20}, as well as the first on-sky deployment of a single-pixel filterbank spectrometer~\cite{deshima20}.
\begin{figure}[b!]
\begin{center}
\includegraphics[width=0.9\textwidth]{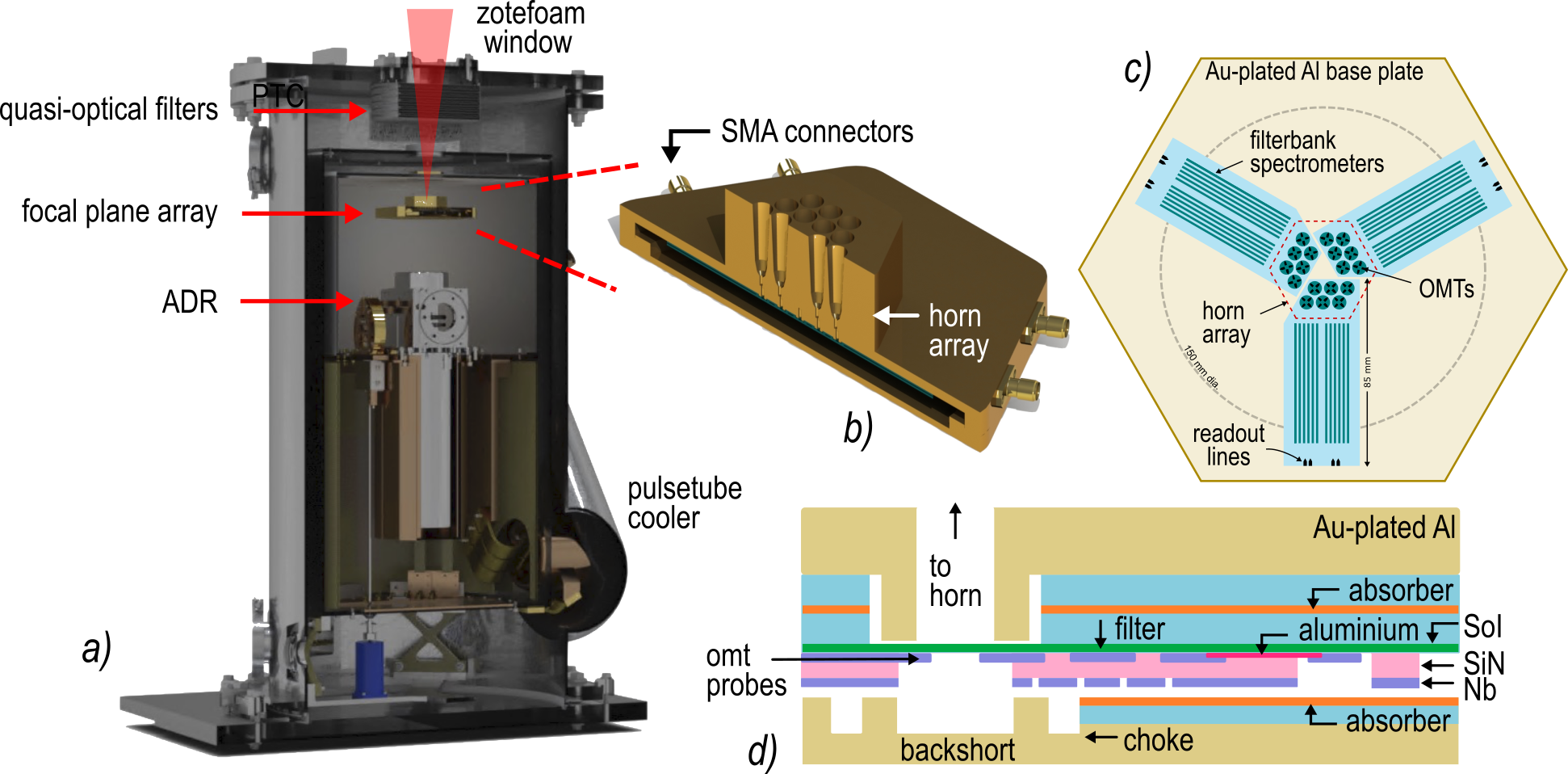}
\caption{ Layout of the SPT-SLIM instrument, {\it a)} a CAD rendering of the SPT-SLIM cryostat, {\it b)} cross section view of the conceptual design of the focal plane array, {\it c)} a schematic top down view of the layout of the sub-module assembly, and {\it d)} cross section view of the dielectric and wafer stack. (Color figure online.) }
\label{fig:fpa}
\end{center}
\end{figure}

The Summertime Line Intensity Mapper is a new pathfinder instrument on the South Pole Telescope (SPT-SLIM) that builds on this technical foundation to demonstrate the in-field performance and suitability of multi-pixel mm-wave filterbank spectrometers and their application to mm-LIM. Combining low-loss mm-wave superconducting circuits with arrays of ultra-sensitive kinetic inductance detectors (KIDs), SPT-SLIM will serve as a key demonstration for large-format highly multiplexed spectroscopic imaging arrays. Scheduled for deployment in 2023, the SPT-SLIM cryostat (Fig.~\ref{fig:fpa}\red{a}) is being constructed as a drop-in replacement for the Event Horizon Telescope VLBI receiver~\cite{marrone19} and will observe during the austral summer. Benefitting from the existing infrastructure at the SPT, a pick-off mirror diverts the telescope beam coming from the primary aperture to a custom-built tertiary mirror that feeds the SLIM focal plane. The optical bandwidth at the detectors is designed to span the 2 mm atmospheric window (120--180 GHz), and is defined through a set of infrared blockers and metal-mesh quasi-optical lowpass edge filters. The focal plane assembly is then cooled to $\sim100$~mK using a cryogen-free adiabatic demagnetisation refrigerator that provides the operating temperature needed to achieve the desired detector sensitivity.

In this paper we present an overview of the SPT-SLIM focal plane and detector architecture, outlining a number of preliminary design choices and considerations. For further details on the scientific motivation, predicted performance of the instrument, as well as a description of other aspects of SPT-SLIM, see Karkare et al.~\cite{karkare21}.

\vspace{-16pt}
\section{Layout of the SPT-SLIM Focal Plane}

A schematic of the SPT-SLIM focal plane is shown in Fig.~\ref{fig:fpa}. The spectrometer array is made up of 18 dual-polarization pixels (36 spectrometers total) arranged in a hexagonal close-packed configuration. The current design is assembled from three separate submodules, where each submodule is fabricated from a single 100 mm diameter wafer containing 12 filterbank spectrometers.

An array of profiled smooth-walled horns (see Fig.~\ref{fig:fpa}\red{b}) is used to provide a well-controlled symmetric beam with low side-lobe level to ensure acceptable image quality in the absence of a cold optical Lyot stop. Radiation is guided onto the microstrip transmission line with a waveguide-coupled planar orthomode transducer (OMT). Orthogonal polarizations feed separate filterbanks, and opposite probes are combined using a 180 degree hybrid coupler that enables single-moded operation over a 2:1 bandwidth with high efficiency~\cite{mcmahon09}, well beyond the 40\% bandwidth needed to achieve the initial SPT-SLIM specifications. The OMT-based antenna design leverages extensive technical heritage from the technology development for cosmic microwave background receivers~\cite{henning12,henderson16}. In addition, the thin-film membrane on which the OMT is patterned serves as an effective method for reducing coupling of stray light to the detectors, which is critical to achieving the strict levels of isolation between adjacent spectrometers that will be needed to achieve high-fidelity low-noise optical spectra.

The submodules are aligned using dowel pins to a common machined gold-plated aluminium baseplate containing the $\lambda/4$ backshorts, readout connections, as well as ancillary structures for heat sinking. As shown in Fig.~\ref{fig:fpa}\red{d}, located on either side of the detector wafer is a custom absorber wafer that contains an impedance matched absorbing structure that acts to eliminate stray radiation both in the form of surface waves~\cite{yates18} within the substrate, as well as any radiation propagating in the vacuum gap. The backside absorber will also form a phonon-absorber that mitigates the effects of cosmic ray impacts\cite{daddabbo14}. The various wafers are stacked together and secured with beryllium-copper spring clamps at the edges as well as in the wafer centers, which has been shown to be an effective strategy for minimizing microphonic noise in large of arrays of kinetic inductance detectors~\cite{muscat20}.
\vspace{-16pt}
\section{Superconducting Filterbank Design}

\begin{figure}[thbp]
\begin{center}
  \includegraphics[width=0.8\textwidth]{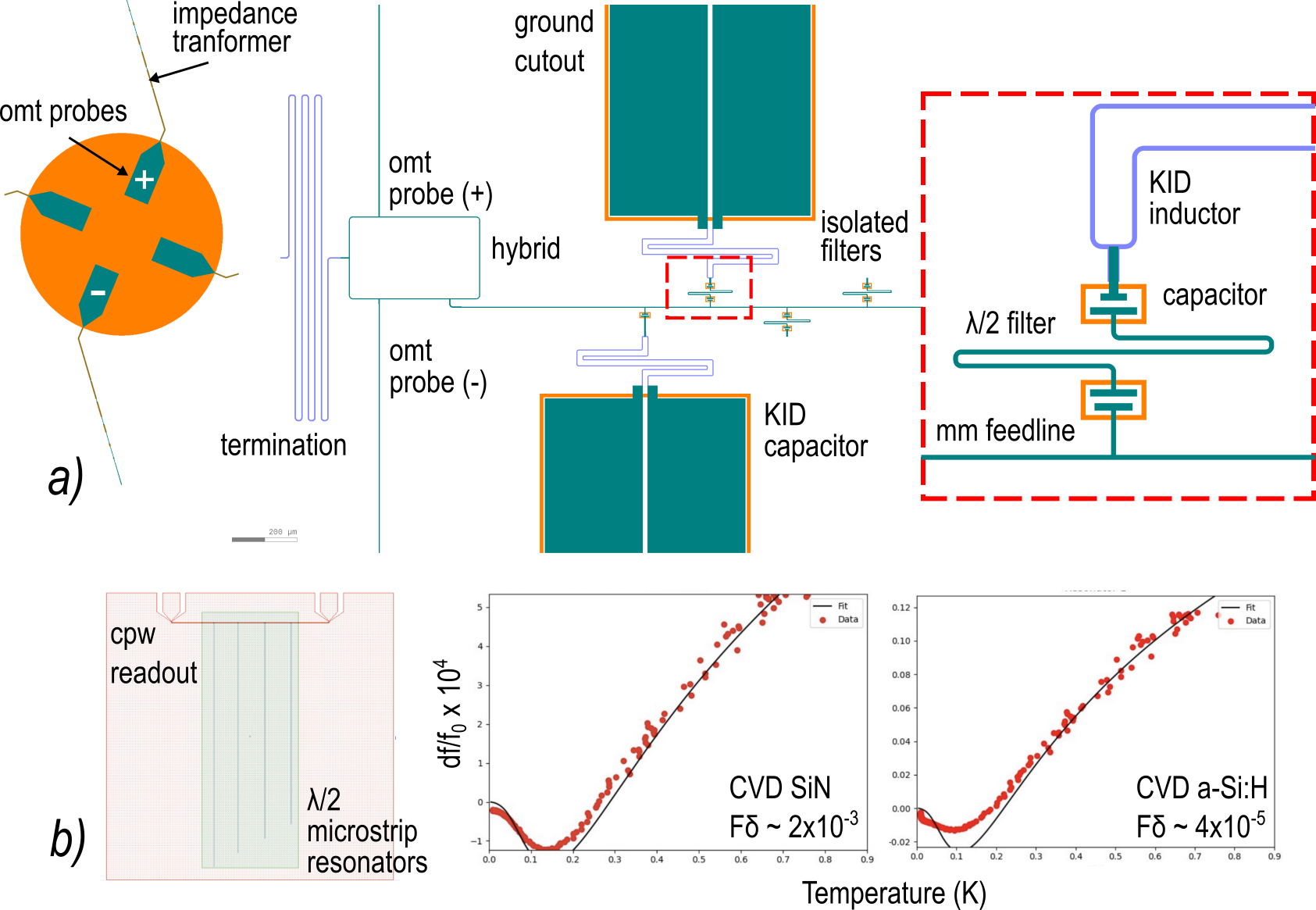}
\caption{ {\it} a) Schematic of a prototype filterbank design and layout, including the OMT probes that are combined an 180-deg hybrid to feed the filter bank. Inset shows a zoom of the micro strip-to-detector coupling. See text for details. {\it b)} Prototype measurements using microwave microstrip resonators (left) to characterise SiN (middle) and a-Si:H (right) (Color figure online).}
\label{fig:filterbank}
\end{center}
\end{figure}

Each SPT-SLIM pixel in the focal plane comprises two $R= \lambda / \Delta \lambda = 300$ superconducting filterbank spectrometers nominally designed to cover the 2 mm atmospheric window. A schematic of a portion of a prototype filterbank layout is shown in Fig.~\ref{fig:filterbank}\red{a}, and is made up of a series of capacitively coupled $\lambda/2$ thin-film microstrip transmission line resonators. The SPT-SLIM filter design has been developed to minimize the effects of fabrication tolerances on the performance, which has been shown to be an issue with existing filter designs~\cite{laguna21}. The input and output capacitors are formed from the series of combination of two lumped-element parallel plate capacitors that use an isolated patch cut into the ground plane. The coupling capacitance sets the bandwidth of the filter and therefore the spectral resolution of each channel. For a detailed description of the simulated performance, optimisation, and tolerance analysis of this filter design see Robson et al.~\cite{robson21}.

The baseline for the SPT-SLIM filterbank design is based on an inverted superconducting microstrip architecture~\cite{shirokoff14}, where the feedline and resonators are deposited on the silicon surface prior to the dielectric and ground plane layers (see Fig.~\ref{fig:fpa}\red{d}). The inverted design allows the detector to be fabricated on a minimally processed surface, which will minimise the contribution from dielectric fluctuations to the detector noise. Futhermore, the photo-sensitive section of the KID resonator is shielded by the ground-plane on the top side, which acts to provide additional protection from stray light, whilst having a negligible effect on the detector noise performance~\cite{hornsby20}.

The material loss tangent of the microstrip dielectric governs the ultimate performance of the spectrometer~\cite{robson21}. Typical amorphous dielectrics are known to exhibit a wide range of values for the dielectric properties that depend strongly on both the specific material, and the particular deposition technique and process conditions. For SPT-SLIM we are exploring several material options in order to achieve the desired spectral resolution whilst maintaining high end-to-end optical efficiency. To screen materials for SPT-SLIM we have adopted a two-step process: performing measurements at microwave frequencies (6-8 GHz) using the temperature dependence of the resonant frequency and loss of a superconducting resonator as a relative probe of the loss tangent (Fig.~\ref{fig:filterbank}\red{b}), and then following up with a characterisation at mm-wave (150 GHz) using on-chip scalar transmission measurements of isolated mm-wave filter channels. For initial prototypes we have baselined CVD silicon nitride (SiN) that has been developed at Argonne~\cite{cecil21}, but are also exploring alternative deposition methods (PVD, PE-CVD), as well as materials that are known to offer improved performance, such as hydrogenated amorphous silicon (a-Si:H). In fact, a preliminary microwave characterisation of CVD-grown a-Si:H suggests a reduction in tan$\delta$ of nearly 50x compared to our baseline SiN process (Fig.~\ref{fig:filterbank}\red{b}). If such an improvement translates to mm-wave, the implementation of a-Si:H for SPT-SLIM could provide a substantial ($>10\%$) increase in per-channel efficiency~\cite{robson21}. Understanding the material performance at mm-wavelengths is an area of on-going study.

\vspace{-16pt}
\section{The microstrip coupled lumped-element KID }

\begin{figure}[ht]
\begin{center}
  \includegraphics[width=0.9\textwidth]{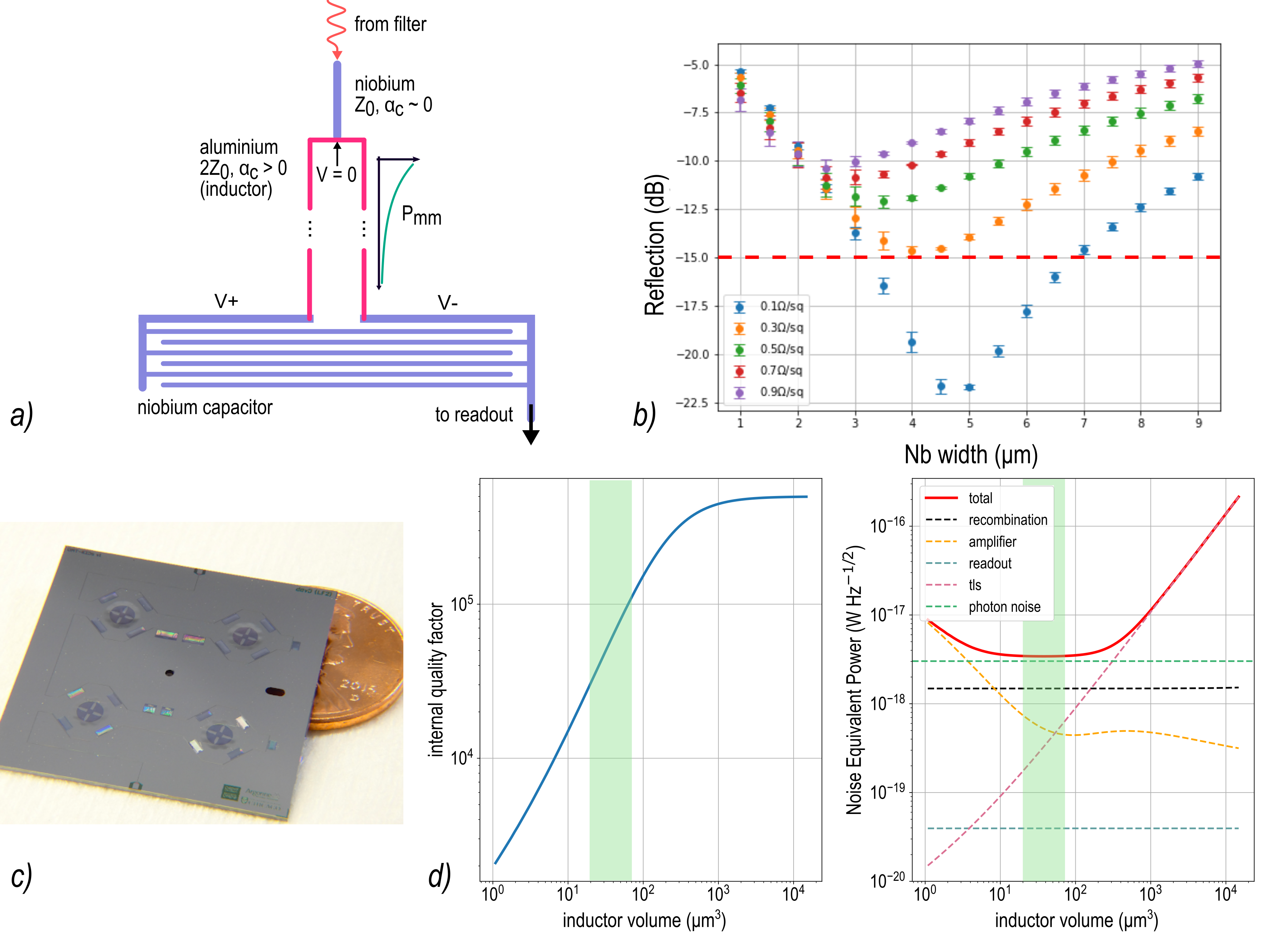}
\caption{  {\it a)} Schematic of the principle of operation of a mc-leKID. {\it b)} Simulated input reflection between the Nb-Al microstrip lines as a function of sheet resistivity. The error bars indicate the peak-to-peak variation across the 120-180 GHz band, {\it c)} Photograph of a preliminary mc-leKID device,  {\it d)} Predicted performance of the mc-leKID as a function of inductor geometry. See text for description. (Color figure online.)}
\label{fig:mclekid}
\end{center}
\end{figure}

Each spectral channel is terminated with a microstrip-coupled lumped-element kinetic inductance detector (mc-leKID)~\cite{tang20}. Providing an elegant and efficient optical coupling to a thin-film microstrip line, the mc-leKID provides a straightforward solution to the highly multiplexed readout for the 10k detectors, and is a key enabling technology for SPT-SLIM. A schematic of the operating principle of mc-leKID is shown in Fig.~\ref{fig:mclekid}\red{a}, and a photograph of a prototype mc-leKID array is shown in Fig.~\ref{fig:mclekid}\red{c}. The output of each mm-wave filter couples to the detector through a microstrip line that feeds the centre (voltage node) of the inductor of the KID resonator, which doubles as a lossy transmission line. As radiation propagates along the KID inductor it is absorbed through the breaking of Cooper pairs and modifies the surface impedance of superconductor. This change is read out as a shift in the resonant frequency, $f_0$, or quality-factor $Q_r$ of the resonator. The inductor must remain a superconducting high-Q material at the KID resonant frequency, whilst absorbing mm-wave photons, which constrains the properties of the superconductor to have a $T_c < 2$ K. For SPT-SLIM, we choose thin-film aluminium. A lumped-element capacitor is used to complete the resonant circuit, and is designed based upon empirical verification of the noise contribution from two-level dielectric fluctuations within the capacitor. We baseline a Nb capacitor fabricated from the same layer as the mm-wave microstrip circuitry to take advantage of the lower TLS noise expected with the Nb-Si interface~\cite{zmuid12} whilst also minimizing the stray-light cross-section and parasitic inductance.

The optimisation of the detector design takes into account the expected optical load, film thickness, and various material properties. The design process for the mc-leKID differs from that from the traditional leKID~\cite{doyle08,mckenney12}, where the inductor is designed to be an efficient free-space absorber, which motivates materials with high normal state resistivity\cite{leduc10}. However, in the case of the mc-leKID the opposite case arises; to minimize reflections at the impedance matched junction the loss in the inductor material should be small in order to minimize the imaginary component of the characteristic impedance ($Z_0$) of the lossy transmission line. Fig.~\ref{fig:mclekid}\red{b} shows the simulated mismatch that arises with increasing the sheet resistivity of a fixed geometry Al inductor/microstrip line. The minimum reflection ($S_{11}$) increases with the film resistivity, and a microstrip geometry of \SI{2}{\um} Al linewidth with 500 nm SiN dielectric thickness requires $R_s <$ \SI{0.3}{\ohm\per\sq} in order to achieve a $<15$ dB return loss (96\% power transfer). As a result, this precludes the use of high-sheet resistance materials $R_s \gg$ \SI{1}{\ohm\per\sq}, and results in an important trade-off between coupling efficiency and detector responsivity.

To study this trade-off, we estimate the detector performance using a model that is based on a simplified calculation of the quasiparticle density~\cite{zmuid12}. By combining the contribution from several pair-breaking mechanisms including optical and readout power, thermal excitations, and an empirically derived residual quasiparticle population, we infer an effective quasiparticle temperature that can differ from the substrate phonon temperature. This effective temperature and total quasiparticle density are then used to calculate the expected resonator parameters ($f_0, Q_i$), detector responsivity, and overall sensitivity. Applying this framework to the mc-leKID, Fig.~\ref{fig:mclekid}\red{d} shows the estimated $Q_i$ and noise equivalent power (NEP) as a function of inductor volume, where the inductor geometry is varied assuming a fixed inductor linewidth (i.e., $Z_0$), while simultaneously varying the film thickness, and subsequently $R_s$, as well the inductor length to ensure a constant level of signal attenuation (-20 dB).

For an $R = 300$ spectral channel under typical atmospheric conditions at the South Pole, we expect an optical loading of around \SI{25}{\femto\watt}, with a corresponding photon noise NEP of $3\times10^{-18}$~W~Hz$^{-1/2}$. Assuming a fixed two-level system noise power spectral density of $S_{\textrm{xx}} \sim 1\times10^{-18}$~Hz$^{-1}$, which is representative of our recent devices~\cite{dibert21}, Fig.~\ref{fig:mclekid}\red{d} shows that in the range of inductor volumes spanning \SIrange[range-units=single]{20}{70}{\cubic\um} the sensitivity is limited by the photon noise and associated recombination noise (green band). A higher $Q_i$ will help relax the multiplexing requirements, and pushes the design to the high end of this range. To reach the multiplexing goal of 2k detectors per readout line, we have also demonstrated the process of lithographically modifying resonators after testing~\cite{shu18,mckenney19}, which enables near-perfect detector yield by modifying the resonant frequencies to separate collided resonators. With our measured ability to place resonators with a fractional accuracy of $2\times10^{-5}$~\cite{mcgeehan18}, combined with the estimated $Q_i$ under load, we expect to achieve extremely high ($>90\%$) operational detector yields for SPT-SLIM.

\vspace{-16pt}
\section{Conclusions}
In this paper, we have presented an overview of various aspects of the design of the SPT-SLIM spectroscopic focal plane and backend detector architecture. The SPT-SLIM focal plane is set to advance superconducting filterbank spectrometer technology and will represent the first in a new class of multi-pixel imaging spectrometers operating a mm-wavelengths. In particular, combining breakthrough detector technology with the world leading observing platform of the SPT, SPT-SLIM is poised to serve as an important scientific and technical demonstration of on-chip spectrometer arrays for a future wide-field mm-LIM surveys.

\begin{acknowledgements}
We acknowledge the efforts of Dave Czaplewski, Margarita Lisovenko, and Volodymyr Yefremenko for the fabrication of the microwave resonator devices. This work is supported by Fermilab under award LDRD-2021-048 and by the National Science Foundation under award AST-2108763. K. S. Karkare is supported by an NSF Astronomy and Astrophysics Postdoctoral Fellowship under award AST-2001802. Work at Argonne, including use of the Center for Nanoscale Materials, an Office of Science user facility, was supported by the U.S. Department of Energy, Office of Science, Office of Basic Energy Sciences and Office of High Energy Physics, under Contract No. DE-AC02-06CH11357.

\end{acknowledgements}

\noindent {\small \textbf{Data availability statement}: The datasets generated during and/or analysed during the current study are available from the corresponding author on reasonable request.}


\end{document}